\documentclass{aa} 
\usepackage{graphicx} 
\usepackage{soul}
\usepackage{txfonts}
\usepackage{xcolor}
\usepackage{tabularx}
\usepackage[]{hyperref}
\usepackage[flushleft]{threeparttable}
\usepackage[switch]{lineno}
\newcommand{\orcid}[1]{\href{https://orcid.org/#1}{\protect\includegraphics[width=8pt]{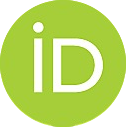}}}

\title{Evolution of star clusters with initial bulk rotation via $N$-body simulations}

\date{---}

\begin{document}
  \author{Abylay Bissekenov\inst{1,2,3}\orcid{0009-0003-4608-2611} \and 
  Xiaoying Pang \fnmsep\thanks{\email{Xiaoying.Pang@xjtlu.edu.cn}}\inst{1,4}\orcid{0000-0003-3389-2263} \and Albrecht Kamlah\inst{5}\orcid{0000-0001-8768-4510} \and
   M.B.N. Kouwenhoven\inst{1}\orcid{0000-0002-1805-0570} \and Rainer Spurzem\inst{5,6,7}\orcid{0000-0003-2264-7203} \and Bekdaulet Shukirgaliyev\inst{8,2,3}\orcid{0000-0002-4601-7065} \and Mirek Giersz \inst{9}\orcid{0000-0002-5987-5077} \and Abbas Askar\inst{9}\orcid{0000-0001-9688-3458} \and Peter Berczik
\inst{9,10,11,12}\orcid{0000-0003-4176-152X} }

   \institute{
            Department of Physics, Xi'an Jiaotong-Liverpool University, 111 Ren'ai Road, Dushu Lake Science and Education Innovation District, Suzhou 215123, Jiangsu Province, P.R. China
            \and
            Energetic Cosmos Laboratory, Nazarbayev University, 53 Kabanbay Batyr Ave, 010000 Astana, Kazakhstan
            \and
            Heriot-Watt University Aktobe Campus, K. Zhubanov Aktobe Regional University, 34 A. Moldagulova Ave, 030000 Aktobe, Kazakhstan
            \and
            Shanghai Key Laboratory for Astrophysics, Shanghai Normal University, 100 Guilin Road, Shanghai 200234, PR China
            \and 
            Astronomisches Rechen-Institut, Zentrum f\"{u}r Astronomie, University of Heidelberg, Mönchhofstrasse 12–14, 69120 Heidelberg, Germany
            \and
            Kavli Institute for Astronomy and Astrophysics, Peking University, Yiheyuan Lu 5, Haidian Qu, 100871 Beijing, China
            \and
            National Astronomical Observatories and Key Laboratory of Computational Astrophysics, Chinese Academy of Sciences, 20A Datun Rd., Chaoyang District, 100101 Beijing, China
            \and
            Department of Physics, School of Sciences and Humanities, Nazarbayev University, 53 Kabanbay Batyr Ave., 010000 Astana, Kazakhstan
            \and 
            Nicolaus Copernicus Astronomical Center, Polish Academy of Sciences, Bartycka 18, 00-716 Warsaw, Poland
            \and
            Main Astronomical Observatory, National Academy of Sciences of Ukraine, 27 Akademika Zabolotnoho St., 03143 Kyiv, Ukraine
            \and
            Fesenkov Astrophysical Institute, 23 Observatory str., 050020 Almaty, Kazakhstan
            \and
            Konkoly Observatory, Research Centre for Astronomy and Earth Sciences, HUN-REN CSFK, MTA Centre of Excellence, Konkoly Thege Mikl\'{o}s \'{u}t 15–17, 1121 Budapest, Hungary
}

\date{Received 10 February 2025 / Accepted 20 May 2025}

\abstract
{Young star clusters can inherit bulk rotation from the molecular clouds from which they have formed. This rotation can affect the long-term evolution of a star cluster and its constituent stellar populations. In this study, we aim to characterize the effects of different degrees of initial rotation on star clusters with primordial binaries. The simulations were performed using NBODY6++GPU. We find that initial rotation strongly affects the early evolution of star clusters. Rapidly rotating clusters show angular momentum transport from the inner parts to the outskirts, resulting in a core collapse. Angular momentum transport is accompanied by a highly elongated bar-like structure morphology. The effects of bulk rotation are reduced on the timescale of two-body relaxation. Rotating and nonrotating clusters experience changes in the direction of angular momentum near the dissolution and early evolution due to the tidal field, respectively. We present synthetic observations of simulated clusters for comparison with future observations in filters of Gaia, CSST, and HST. This work shows the effects of bulk rotation on systems with primordial binaries and could be used for the identification of rotation signatures in observed open clusters.}

    \titlerunning{Evolution of star clusters with initial bulk rotation}
   \authorrunning{Bissekenov et al.}

\maketitle

%
\section{Introduction}

The dynamics of stellar systems is a major research field of astrophysics. It focuses on the gravitational interactions of stars in star clusters and can provide crucial insights into stellar evolution and Galaxy formation \citep[see, e.g.,][]{chandrasekhar2005principles,  binney2011galactic, krumholz_star_2019}. It is also essential for the study of the constituents of stellar systems, such as binaries, black holes, and neutron stars, and to the characterization of their dynamical and observational effects on dense stellar systems \citep[see, e.g.,][]{sridhar1999, ivanova2006formation, alexander2017stellar, IBP2024, BPI2025}. The study of stellar dynamics can provide constraints on the conditions of star formation and can be used to examine the effects of properties on the system. One of such properties is bulk rotation, which is the rotational motion of a star cluster around its center of mass. Star clusters inherit their initial rotation from molecular clouds through the star formation process \citep{mapelli2017, lahen2020mar}. At later times, bulk rotation can evolve through two-body relaxation, tidal interactions with other star clusters or molecular clouds, and tidal interactions with the Galactic tides. Over time, bulk rotation can significantly influence the morphology, core collapse process, and long-term evolution of star clusters.

Numerous observations have confirmed the presence of bulk rotation in star clusters. \cite{bianchini2018} observed 81 globular clusters and found clear evidence of rotation in 21 of them. Furthermore, \cite{pang2021a} found velocity anisotropy among the elongated clusters. They suggest that velocity anisotropy might be induced by global rotation. There is also evidence that globular clusters in \texttt{Gaia} Data Release 3, NGC\,104 and NGC\,2904, show clear evidence of a rotation-mass relation \citep{scalco2023}. \cite{jadhav2024} observed the proper motions and radial velocities of 1379 open clusters and identified rotation in ten clusters, and suggested 16 cluster candidates with rotation. \cite{hao2024} discovered a 3D rotation signature in three open clusters, Pleiades, Alpha Persei, and Hyades, using Gaia\,DR3 proper motions and radial velocities.

Initial bulk rotation has been numerically studied using hydrodynamical simulations of star formation in molecular clouds. \cite{mapelli2017} found that initial bulk rotation originates from molecular clouds as a result of interactions with gas filaments and clumps. They indicated that rotation should be commonly noticeable in massive embedded star clusters with masses of $M\approx 4\times 10^3\,M_\odot$. \cite{lahen2020nov} investigated star clusters formed in metal-poor starburst dwarf galaxies and spotted significant angular momentum upon formation in their simulations. They found that more massive clusters tend to have a greater specific angular momentum, and that the angular momentum is not always correlated with flattening.  \cite{ballone2020}  further confirmed that simulated clusters initialized with significant rotation develop substructuring and show a high degree of fractality.

The dynamical evolution of rotating stellar systems can also be studied using 2D orbit-averaged Fokker-Planck (FP) models \citep{einsel1999spurzem}. FP models are extensions of the King model \citep{kingmodel} and based on the numerical solution of the orbit-averaged FP equation \citep{goodman1983, lupton1987, longaretti1996}. These models explained the influence of rotation and its effects on the dynamical evolution of star clusters. Further works based on FP models include studies on binary heating and a stellar mass spectrum \citep{kim2002einsel,kim2004lee, kim2008yoon, fiestas2006}. \citet{fiestas2006} stated that globular clusters in the Milky Way were initially flattened due to rotation and that their current flattening can also be explained by bulk rotation. This also suggests that open clusters may form with a certain amount of bulk rotation.

In addition, $N$-body simulations have also been used to investigate rotating star clusters. \cite{ernst2007}, \cite{kim2008yoon}, and \cite{hong2013} used direct $N$-body simulations to evaluate the accuracy of the numerical solutions of the FP model. They generally found a fair agreement between both models in all three phases of cluster evolution (core collapse, post-collapse, final dissolution). Rotation in nuclear star clusters has been studied with $N$-body models and FP models in \cite{fiestas2010} and \cite{fiestas2012}, although their main focus was the evolution of black holes. The $N$-body models of \citet{tiongco2022} and \citet{livernois2021, livernois2022} were all carried out with a larger particle number ($N=50\,000$) than earlier studies. These studies, however, initialized their models with a limited stellar initial mass spectrum (only stars with masses less than one solar mass). \cite{kamlah22} performed $N$-body simulations of rotating star clusters with and without stellar evolution. In comparison to previous works, \citet{kamlah22} used a more realistic initial mass function with stellar masses up to $150\,M_\odot$ and $1.1\times10^5$ particles. They examined the effects of different initial bulk rotation speeds on systems with an initial binary fraction of 10\%. In their analysis, they focused on the impact of the coupled phenomenon of the gravothermal-gravogyro catastrophe and its relation to rotating systems.

Despite the significant efforts to study the impact of bulk rotation on the evolution of star clusters and their constituents, there are still areas where understanding can be improved by further theoretical analysis. Hydrodynamical simulations are the most detailed and realistic, but they are computationally expensive and only relevant during the early stages of star cluster evolution. FP simulations and $N$-body simulations, on the other hand, produce similar results for the models with moderate bulk rotation, but have slight differences in models with high rotation \citep{hong2013}. Several of the $N$-body studies have used an equal-mass spectrum or an initial mass function with stellar masses below 1\,$M_{\odot}$ mass, which affects the overall dynamics \citep{tiongco2018, tiongco2022, livernois2021, livernois2022, scalco2023}. With the exception of \cite{kamlah22}, previous studies do not account for the presence of primordial binaries. The impact of binary systems is particularly important, considering the gravothermal-gravogyro catastrophe that leads to contraction and core collapse. Binaries can delay the contraction and affect the dynamical evolution of rotating systems. Evaluating the impact of binarity is therefore important for a better understanding of the evolution of rotating star clusters. Finally, earlier numerical studies tend not to compare the outcomes of the simulations with observations (e.g., absolute magnitudes and spectral energy distributions) which is the final objective of theoretical studies such as these.

In this paper, we investigate the dynamical evolution of star clusters with different primordial binary fractions, metallicities, and initial bulk rotations. We identify the rotation signature and discuss the effect on the dynamical evolution of the star clusters. This work highlights the impact of primordial binaries and metallicities on star cluster dynamics and the evolution of the 3D morphology of star clusters. Additionally, we present our data in the form of synthetic observations for future comparison with observations.

The paper is organized as follows. In Section~\ref{sec:nbody} we describe the initial conditions and the $N$-body simulations. In Section~\ref{sec:results}, we present and discuss the results in terms of rotation signatures (Section~\ref{sec:ang_mom}), dynamical evolution (Section \ref{sec:bin}), morphology (Section~\ref{sec:morphology}), and synthetic observations (Section~\ref{sec:galevnb}). In Section \ref{sec:dis}, we analyze the star cluster properties simultaneously and discuss their correlations. Finally, in Section~\ref{sec:conc}, we summarize and conclude the findings of our work. 


\section{Methodology}
\label{sec:nbody}

We simulated 20 star cluster models using \texttt{NBODY6++GPU}, a high-performance GPU-accelerated code for gravitational $N$-body simulations \citep{kamlahnbody}. \texttt{NBODY6++GPU} is a successor of the direct force integration $N$-body codes originally written by Sverre Aarseth \citep{aarseth1985, aarseth1999, aarseth2003, spurzem1999, aarsethtoutmardling2008}.  Using these direct $N$-body simulations, we studied the impact of initial rotation, stellar evolution, binary fraction, and the combination of catastrophes. Our models use \texttt{level C} stellar evolution in the form presented by \cite{kamlahnbody}. 

The models were initialized using \texttt{McLuster} \citep{kupper2011, kamlahnbody, leveque2022}.  \texttt{McLuster} can set up initial conditions of $N$-body simulations or generate star clusters from observations \citep{kupper2011}.

We used \textsc{FOPAX} \citep{einsel1999spurzem, kim2002einsel, kim2004lee, kim2008yoon} to generate the 2D FP initial models that add rotation to initialized $N$-body simulations, following a similar approach as in \cite{einsel1999spurzem} and \citep{hong2013}. The code produces a 2D mesh from the output of density $\rho$ and velocity dispersions $\sigma$ as a function of $r$ and $z$ of a rotating King model $f(E, J_z)$. The density model is based on the King model \citep{kingmodel} that can be expressed as
\begin{equation}
    f_{rk} \propto (e^{\beta E}-1)\times e^{-\beta \Omega_0 J_z}
    \quad .
\end{equation}
Here, $\beta=(m\sigma^2)^{-1}$ is the anisotropy parameter (where $\sigma$ is the velocity dispersion and $m$ is the stellar mass), $\omega_0=\Omega_0\sqrt{9 \pi Gn_c/2}$ is the angular velocity, and $W_0=\beta m(\psi-\psi_t)$ is the King parameter, in which $\psi$ and $\psi_t$ are King potentials at the center and at the truncation radius, respectively \citep{lupton1987, einsel1999spurzem}. King models are usually described by the pair of parameters  $(W_0, \omega_0)$ \citep{einsel1999spurzem}. For our models, we adopted $W_0=6.0$ and $\omega_0 \in [0.0, 0.6, 1.2, 1.8]$. 

We used a Monte Carlo rejection technique to generate the discrete system of $N$ particles that follow the distributions of $r$ and $z$. All the output is in $N$-body format (one line per particle, containing the mass, 3D position, and  3D velocity data). We combined the generated $N$-body velocity distribution with the position distribution generated by \textsc{McLuster} (see above). All the data were then scaled to standard H\'enon units. Binary orbital parameter distributions were retained from \textsc{McLuster}. Finally, we obtained the rotating King model with binary systems and the \cite{kroupa} initial mass function.

Initial conditions of the simulations are summarized in Tab.~\ref{tab:description}. The initial total number of single stars and binary systems in each system is $n=N_{\rm single}+N_{\rm bin} = 10\,000$. We adopted a binary fraction $f_{\rm bin}$, which is the ratio of the cluster members that are binary systems: 
 
\begin{equation}
    f_{\rm bin}=\frac{N_{\rm bin}}{N_{\rm single}+N_{\rm bin}} \times 100\%
    \quad .
\end{equation}

Here, $N_{\rm bin}$ and $N_{\rm single}$ are the number of binary systems and single stars, respectively. We modeled three binary fractions $f_{\rm bin}=0\%$, 10\%, and 50\%, such that the total number of stars in these models is $N=N_{\rm single}+2N_{\rm bin}=10\,000$, 11\,000, and 15\,000, respectively. The corresponding initial total cluster masses are $5.8\times10^3\,M_\odot$, $6.4\times10^3\,M_\odot$, and $8.79\times10^3\,M_\odot$. Binary systems are generated using a uniform mass ratio distribution ($0.1 \leq q=m_2/m_1 \leq 1.0$) for $m_1>5$\,$M_\odot$, and random pairing is adopted for the remaining binary systems. The semimajor axes are distributed uniformly in $\log a$ between the sum of the two stellar radii and 100\,au. We adopted a thermal eccentricity distribution, $f(e)=2e$. We adopted a \cite{kroupa} initial mass function with stellar masses between 0.08\,$M_\odot$ and 150.0\,$M_\odot$. For each of these models, we studied two different metallicities: metal-poor clusters ($Z=0.01$) and clusters with solar metallicity ($Z=0.02$). Finally, we modeled four degrees of bulk rotation $\omega_0=0.0$, 0.6, 1.2, and 1.8.

\begin{table}[tb]
    \centering
    \scriptsize
    \caption{Initial conditions and parameters for the models.}
    \label{tab:model_params}
    \begin{tabular}{c c c c c c c} 
        \hline
        \textbf{Model} & \textbf{$N$} & \textbf{$f_{\rm bin}$} & \textbf{$Z$} & \textbf{$\omega_0$} & $r_h$ [pc]  & $r_t$ [pc] \\
        \hline
        N10k\_fb00\_z001\_w000 & $10\,000$ & 0\%& 0.01 & 0.0 & 2.25 & 21.65\\ 
        N10k\_fb00\_z001\_w006 & $10\,000$  & 0\%& 0.01 & 0.6 & 2.26 &21.65 \\
        N10k\_fb00\_z001\_w012 & $10\,000$  & 0\%& 0.01  & 1.2 & 2.23 & 21.65 \\
        N10k\_fb00\_z001\_w018 & $10\,000$  & 0\% & 0.01& 1.8 & 2.45 & 21.65 \\
        \hline
        N11k\_fb01\_z001\_w000 & $11\,000$ & 10\% & 0.01 & 0.0 & 2.03 & 22.38 \\
        N11k\_fb01\_z001\_w006 & $11\,000$  & 10\%  & 0.01 & 0.6 & 2.18 & 22.38 \\
        N11k\_fb01\_z001\_w012 &$11\,000$  & 10\% & 0.01 & 1.2 & 2.28 & 22.38 \\
        N11k\_fb01\_z001\_w018 &$11\,000$  & 10\%  &0.01 & 1.8 & 2.40 & 22.38 \\
         \hline
        N11k\_fb01\_z002\_w000 & $11\,000$  & 10\% & 0.02 & 0.0 & 2.03 & 22.38 \\
        N11k\_fb01\_z002\_w006 & $11\,000$  & 10\%  & 0.02 & 0.6 & 2.18 & 22.38 \\
        N11k\_fb01\_z002\_w012 & $11\,000$  & 10\%   & 0.02 & 1.2 & 2.28 & 22.38 \\
        N11k\_fb01\_z002\_w018 & $11\,000$  & 10\%  & 0.02 & 1.8 & 2.41 & 22.38 \\
        \hline
        N15k\_fb05\_z001\_w000 & $15\,000$ & 50\% & 0.01 & 0.0 & 2.09 & 24.82 \\
        N15k\_fb05\_z001\_w006 & $15\,000$ & 50\% & 0.01 & 0.6 & 2.27 & 24.82 \\
        N15k\_fb05\_z001\_w012 & $15\,000$ & 50\% & 0.01 & 1.2 & 2.36 & 24.82 \\
        N15k\_fb05\_z001\_w018 & $15\,000$ & 50\% & 0.01 & 1.8 & 2.41 & 24.82 \\
        \hline
        N15k\_fb05\_z002\_w000 & $15\,000$ & 50\% & 0.02 & 0.0 & 2.09 & 24.82 \\
        N15k\_fb05\_z002\_w006 & $15\,000$ & 50\% & 0.02 & 0.6 & 2.27 & 24.82 \\
        N15k\_fb05\_z002\_w012 & $15\,000$ & 50\% & 0.01 & 1.2 & 2.36 & 24.82 \\
        N15k\_fb05\_z002\_w018 & $15\,000$ & 50\% & 0.02 & 1.8 & 2.41 & 24.82 \\
        \hline
    \end{tabular}
    \begin{tablenotes}
      \small
      \item \textbf{Notes.} Columns list the model name, the total number of initial stars, $N$, the binary fraction, $f_{\rm bin}$, the metallicity, $Z$, the initial rotation parameter, $\omega_0$, and half mass radius, $r_h$.
    \end{tablenotes}
    \label{tab:description}
\end{table}

We initialized the models with $N=10\,000$, 11\,000, and 15\,000 stars with half-mass radii of $r_h=2.25$, 2.03, and 2.09\,pc, respectively.  The initial distributions of King models become more concentrated for larger values of $\omega_0$  (see Fig. 1 in \citet{einsel1999spurzem}). After initializing the rotating clusters with the input parameters of the King model with \textsc{McLuster},  structural parameters from \textsc{McLuster} such as $r_h$ are slightly changed. Thus, the initial half-mass radii of the model realizations are $r_{h}\approx2-2.5$ depending on the choice of $\omega_0$ (the second column from the left in Tab.~\ref{tab:description}). The reader could notice that $r_h$ of models with $f_{bin}=0\%$ are larger compared to those with binary fractions. This is because clusters with binaries tend to be smaller compared to those without. However, when the clusters rotate very fast, the binding effect of binaries wears off. Thus, $r_h$ is larger in the modes with high $\omega_0$. This affects clusters with $f_{bin}=0\%$ too, but it is mostly noticeable in the model with $\omega_0=1.8$. The stochastic nature of the initial mass function sampling and random selection of particle coordinates also contribute to an average 0.1\,pc deviation in $r_h$. The tidal field is modeled as a point mass galaxy with a mass $\approx1.4\times10^{11}\,M_\odot$, and the clusters are placed at a circular orbit with a distance $\approx8.178$\,\text{kpc} from the Galactic center.  The models with 10k, 11k, and 15k stars have initial tidal radii of 21.65\,pc, 22.37\,pc, and 24.81\,pc, respectively. We evolved all models until dissolution, which typically occurs after  $1-2.5$\,Gyr.

Particles are removed from the simulation and labeled as escapers once they reach a distance equal to twice the tidal radius at that time. These escapers are excluded from the calculations of any of the star cluster global parameters, such as $r_t$ or $m_c$.


\section{Results}
\label{sec:results}

\subsection{Rotation signatures}
\label{sec:ang_mom}

\subsubsection{Angular momentum}
\label{sec:spec_ang}

We quantify the rotation signature using the total angular momentum vector of the star cluster, $\mathbf{L}_{\text{total}} = \sum_i M_i (\mathbf{r}_i \times \mathbf{V}_i)$. For single stars, $M_i$ is the mass, $\mathbf{r}_i$ is the position relative to the stellar density center of the cluster, and $\mathbf{V}_i$ is the velocity, relative to the stellar density center of the cluster. For binary systems, $M_i$ is the sum of the two stellar masses, and $\mathbf{r}_i$ and $\mathbf{V}_i$ correspond to position and velocity relative to the center of mass of each binary system. The center of mass is adjusted after every energy check in the simulation. We also calculate the angular momentum of the single stars and binary systems within different Lagrangian radii, $\phi$, and examine the angular momentum of the inmost ($\phi=0.0-0.3$), intermediate ($\phi=0.3-0.6$), and outmost ($\phi=0.6-1.0$) regions of the system. We note that we used the total cluster mass at the time in the calculations of Lagrangian radii.

\begin{figure*}[htbp]
    \centering
    \includegraphics[width=\textwidth]{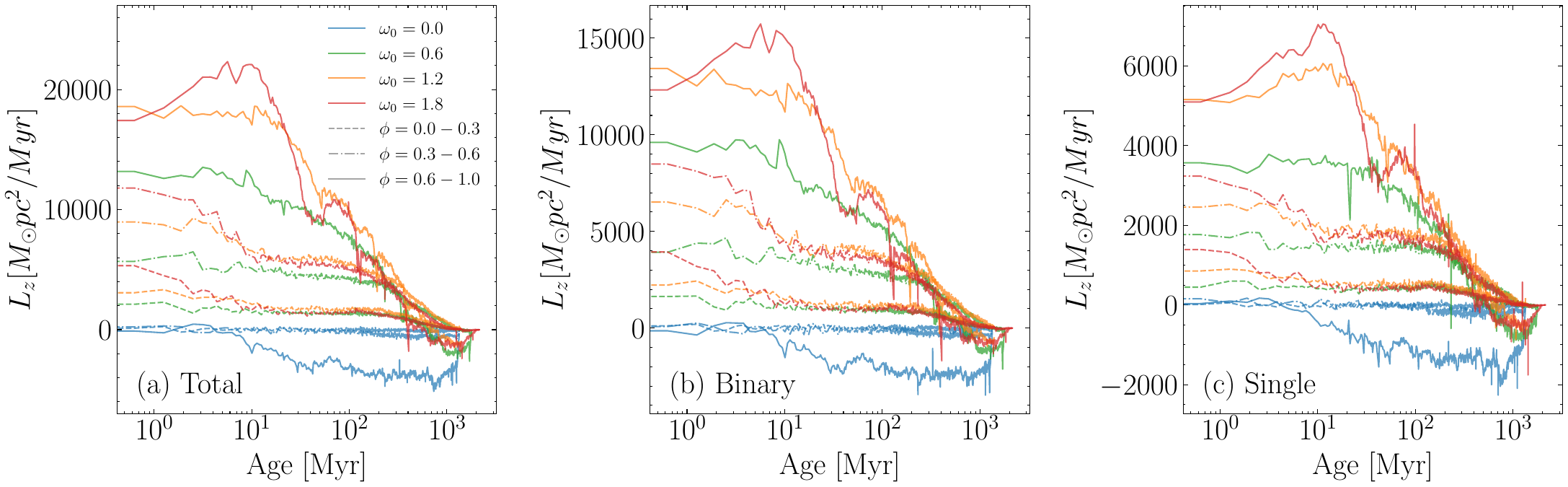}
    \caption{Angular momentum of all stars (panel (a)), binary stars (panel (b)), and single stars (panel (c)) of the simulations with 15k stars, $f_{\rm bin}=50\%$, and $Z=0.01$. Colors represent simulations with different $\omega_0$. The line styles represent the angular momentum of stars within the inmost,  intermediate, and outmost regions.}
    \label{fig:ang_mom}
\end{figure*}

Fig.~\ref{fig:ang_mom} shows the angular momentum of all stars (panel (a)), of the binary stars (panel (b)), and of the single stars (panel (c)), for the simulation with 15k stars, $f_{\rm bin}=50\%$, and $Z=0.01$ as an example. The angular momentum is computed considering the $z$-axis as the rotation axis. We note that only stars $M_{\rm ZAMS}<6$\,$M_\odot$ were used for the calculation of the angular momentum of the total system, the population of binaries, and the population of single stars. This is the optimal subset of stars for calculations since $M_{\rm ZAMS}\geq6$\,$M_\odot$ stars undergo supernova explosions due to the stellar evolution (further explanations are in Sec.~\ref{sec:esc}).

Panel (a) of Fig.~\ref{fig:ang_mom} shows the total angular momentum of the the models as a function of time. All the parts of the rotating systems experience moderate changes up until 10\,\text{Myr}. The model without bulk rotation has angular momentum close to zero in all its parts, and changes in the outer part to a negative direction after $7$\,\text{Myr}. The angular momentum of the inmost parts of all models is seen to be decreasing, and the rate of decrease is higher for models with more bulk rotation. The angular momentum of the intermediate parts of the systems is also seen to be decreasing, but the model with $\omega_0=0.6$ shows some spike increase at $\approx1.5$\,\text{Myr}. There is no significant change in the angular momentum for the outmost parts of all models, except for the model with $\omega_0=1.8$. Such changes of the angular momentum for different parts of the cluster indicate that angular momentum is transported from the inner region to the outer parts of the cluster. Thus, we observed a decrease in the angular momentum of the inmost and intermediate parts. The outmost part receives the momentum from the inner and intermediate parts, but it can be clearly shown only in the model with $\omega_0=1.8$. The reason for this is that models with high rotation tend to have larger core mass and core radius (see Fig.~\ref{fig:4params}). Thus, the retention of stars in outmost parts of models with $\omega_0=[0.6,1.2]$ is lower. Smaller retention of stars does not allow clusters with $\omega_0=[0.6,1.2]$ to hold the angular momentum in the outmost part for long, and we did not see an increase as in the model with $\omega_0=1.8$. The process of angular momentum transport results in a core collapse that can be seen in the decrease of core radius in Fig.~\ref{fig:4params}. This is a gravogyro catastrophe described in the work of \citep{hachisu79}. After 10\,\text{Myr}, the angular momentum of the inmost and intermediate parts reaches the constant values that they will hold on until $\approx400$\,\text{Myr}. The angular momentum of outmost parts in the models is decreasing. This is further followed by the slight increase in momentum at $\approx45$\,\text{Myr} in the models with $\omega_0=[1.2,1.8]$. Further, the outmost part shows a constant decrease and reaches even negative values, and the inmost and central parts reach $0\,\text{pc}^2 \,\text{Myr}^{-1}$ by the time of dissolution. A similar decrease in angular momentum was seen in the work of \citet{livernois2022}. However, negative angular momentum is not seen in the study, since the runtime of their simulations is until the core collapse.

The panels in rows (b) and (c) of Fig.~\ref{fig:ang_mom}  show the angular momentum of the single stars and binary systems of the models. Trends seen in total angular momentum are also seen in the angular momentum of binaries and singles. Angular momentum of binaries is a larger contributor to the total angular momentum compared to single stars. This is mostly because binaries contain twice the mass of single stars in this model with $f_{bin}=50\%$.

\begin{figure*}[tb]
    \includegraphics[width=\textwidth]{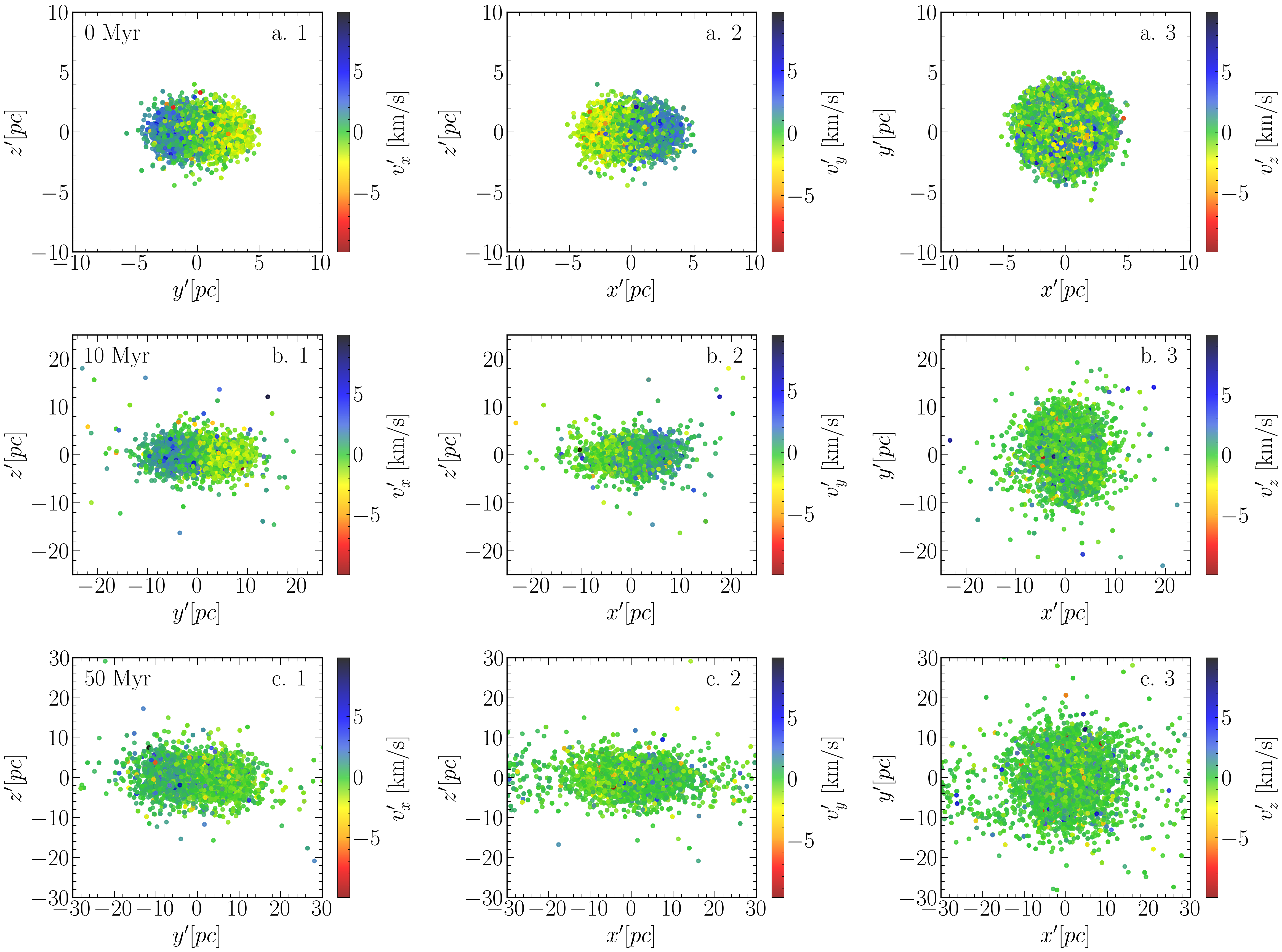}
    \caption{Spatial coordinates for the model with $\omega_0=1.8$ at 0\,Myr (panels (a)), 10\,Myr (panels (b)), and 50\,Myr (panels (c)). The color map shows the component of the stellar velocity along the $x'$-, $y'$-, and $z'$-axes (where the $z'$ axis is the direction of the net angular momentum). }
    \label{fig:velocity}
\end{figure*}

We transformed the coordinates into the angular momentum vector's direction at each time and inspect stars by their velocities such that the net angular momentum of the star cluster points in the direction of the $z'$-axis. Fig.~\ref{fig:velocity} presents the transformed spatial coordinates for models with $\omega_0=1.8$ at 0\,Myr (panels (a)), 10\,Myr (panels (b)), and 50\,Myr (panels (c)). The color map shows the velocity of the stars towards the transformed direction. We note that we limit displayed velocities from $-10$\,km\,s$^{-1}$ to $10$\,km\,s$^{-1}$ to lessen the impact of escapers. From this, we can see that the model with $\omega_0=1.8$ initially has a noticeable rotation signature at $0\,\text{Myr}$ and $10,\text{Myr}$. $v_x'$ and $v_y'$ are seen in two distinctive regions with positive and negative velocities. As for the $v_z'$, it does not display similar positive and negative regions. This is because the $z'$-axis is still the rotation axis. These results are consistent with the hydrodynamical simulations of \citet{mapelli2017} and \citet{ballone2020}, who show similar results but for clusters from a simulated molecular cloud at $3\,\text{Myr}$ and $5\,\text{Myr}$. However, we can also show the velocities at the later stages of evolution. At 50\,Myr, it is seen that the regions are not as distinguishable, and the velocity differs in the range of $-2$\,km\,s$^{-1}$ to 2\,km\,s$^{-1}$. This is due to the reduction of the rotation signature over time, and it continues to decrease further. The two-body relaxation timescale in the models is $\approx96-119$\,\text{Myr}, with clusters with a higher $\omega_0$ having a longer relaxation time. The significant reduction in rotation signature by 50\,Myr and further reduction at later times suggests that the effects of rotation are reduced when the clusters reach one relaxation time. 
The changes of the rotation signature from a few to 0.1\,km/s are consistent with the current observation of open clusters by \citet{hao2024}, who measured the internal rotation of open clusters at the level of 0.1\,km/s. 

\subsubsection{Escapers}
\label{sec:esc}
 
In order to interpret the rotation signature, it is crucial to investigate the impact of escapers on the stellar system. As a star cluster evolves, angular momentum is transported from the inner regions to the outer regions of the cluster and is subsequently removed from the system through escaping stars and binary systems.

Aside from the escapers that stop being gravitationally bound to a system, the stellar evolution of the stars also affects the evolution of the system. In our analysis, we divide the stars into four groups by the zero-age main-sequence mass of stars (ZAMS): $M_{\rm vlm}$. $M_{\rm lm}$, $M_{\rm mm}$, and $M_{\rm hm}$, which correspond to the following mass ranges:
\begin{align*}
    M_{\rm vlm} &: 0.08\,M_{\odot} \leq m_{\rm ZAMS} < 0.9\,M_{\odot} \\
    M_{\rm lm}  &: 0.9\,M_{\odot} \leq m_{\rm ZAMS} < 6\,M_{\odot} \\
    M_{\rm mm}  &: 6\,M_{\odot} \leq m_{\rm ZAMS} < 15\,M_{\odot} \\
    M_{\rm hm}  &: 15\,M_{\odot} \leq m_{\rm ZAMS} < 150\,M_{\odot}
\end{align*}
Here, $m_{\rm ZAMS}$ is the ZAMS mass of a single star. The mass ranges are chosen to separate stars with different evolutionary tracks. Stars with masses $M_{\rm vlm}$ will remain main-sequence stars throughout the entire simulation. Most of those with $M_{\rm lm}$ will become white dwarfs (WDs), those with $M_{\rm mm}$ become neutron stars, and those with $M_{\rm hm}$ become black holes. However, we note that these definitions are not strict, and exceptions do exist. 

\begin{figure}[tb]
    \centering
    \includegraphics[width=0.8\linewidth]{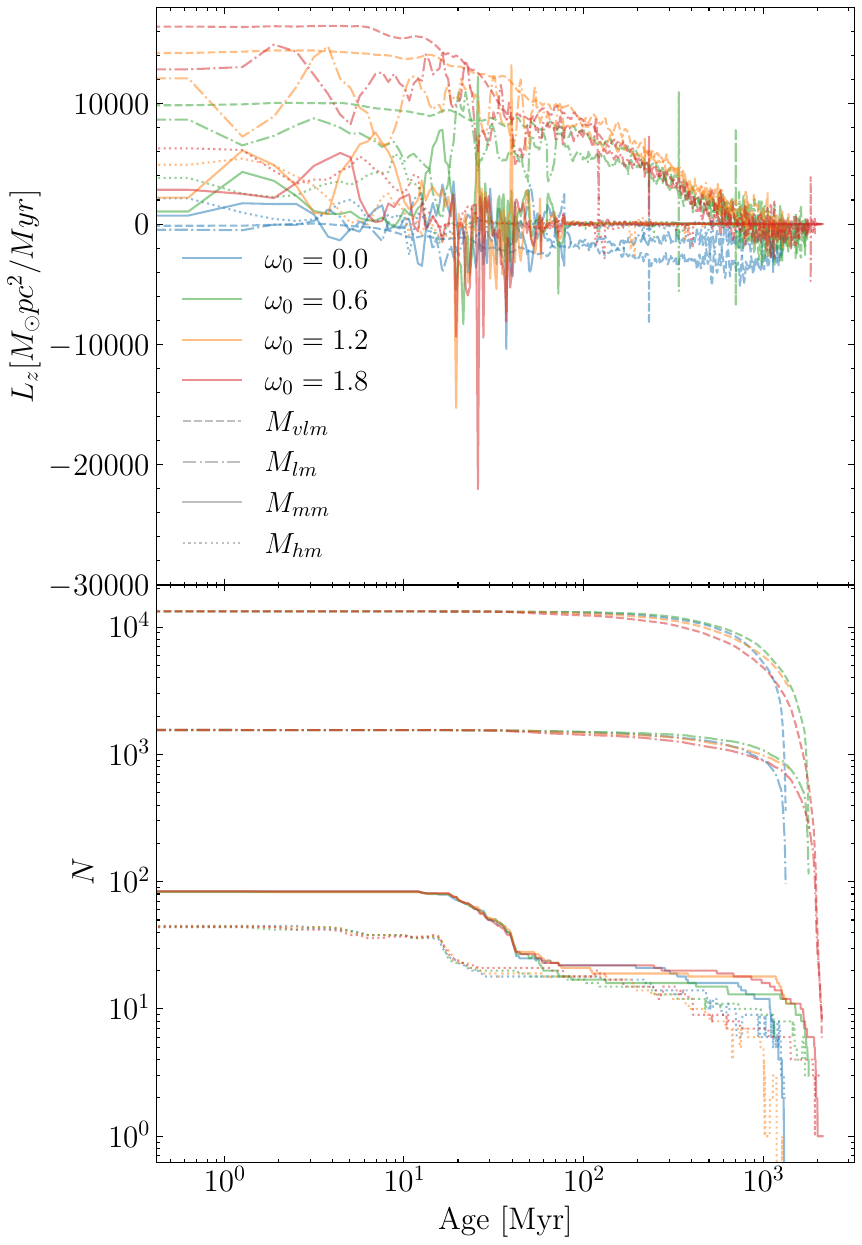}
    \caption{Evolution of the angular momentum (top panel) and number of stars (bottom panel) of four groups, for the simulations with 15k stars, $f_{\rm bin}=50\%$, and $Z=0.01$. Colors scheme as in Fig. \ref{fig:ang_mom}. The line styles represent groups of stars by $M_{\rm ZAMS}$.}
    \label{fig:groups}
\end{figure}

Fig.~\ref{fig:groups} shows the sum of the angular momentum (top panel) and number of stars (bottom panel) of the four groups of stars as a function of time of the simulations with 15k stars, $f_{\rm bin}=50\%$, and $Z=0.01$. The $M_{\rm vlm}$ stars follow the global trend of the angular momentum of Fig.~\ref{fig:ang_mom}. This is the largest group of stars in the cluster, as seen in the bottom panel of Fig.~\ref{fig:groups}. The 1500 stars in the $M_{\rm lm}$ group show some fluctuations in the angular momentum,  but the trend of angular momentum is unaffected. For the stars in the $M_{\rm mm}$ and $M_{\rm hm}$ groups, there are strong fluctuations around $20-40$\,\text{Myr}. Considering the comparatively low number of stars (less than 100) in the groups $M_{\rm mm}$ and $M_{\rm hm}$, the statistical fluctuations are considerably high. The number of stars in these two groups suffers from a sharp decrease during this period, which is due to stellar evolution. These stars undergo supernova explosions, that lead to observed extreme values in angular momentum (spikes in the upper panel). These fluctuations might bias our analysis of the angular momentum of the cluster. Therefore, we excluded stars with $M_{\rm ZAMS}\geq6$\,$M_\odot$ from the angular momentum calculations from Sec. \ref{sec:ang_mom}. 

\begin{figure}[tb]
    \centering
    \includegraphics[width=0.8\linewidth]{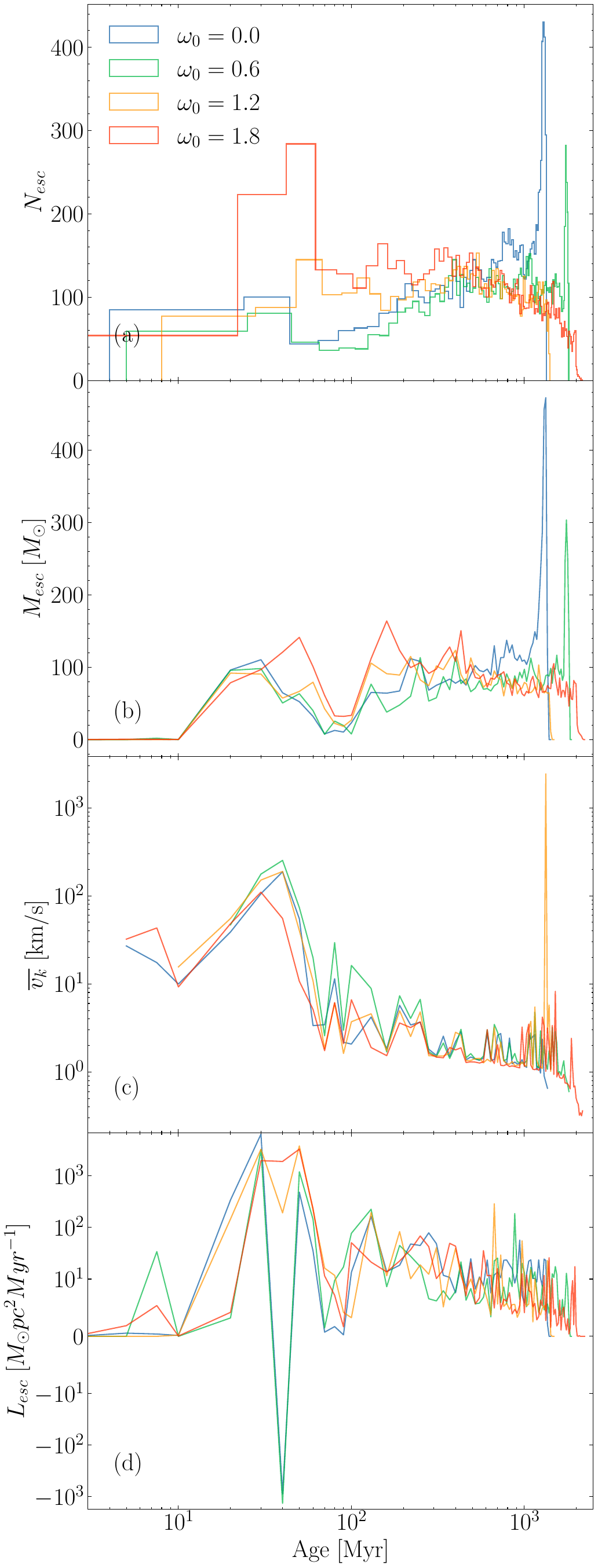}
    \caption{Properties of escaping stars over time: Evolution of the number of stars $N_{\rm esc}$ (panel (a)), total mass $M_{\rm esc}$ (panel (b)), mean kick velocity $\overline{v}_k$ (panel (c)), and angular momentum $L_{\rm esc}$ (panel (d)). The color scheme is the same as that of Fig.~\ref{fig:ang_mom}.  }
    \label{fig:esc}
\end{figure}

Fig.~\ref{fig:esc} shows the properties of escaping stars such as the number of stars $N_{\rm esc}$ (panel (a)), the total mass of escapers $M_{\rm esc}$ (panel (b)), mean kick velocity $v_{k}$ (panel (c)), and angular momentum $L_{\rm esc}$ (panel (d)) as a function of time. Unlike in previous calculations, the total angular momentum of escapers is calculated by $L_{\rm esc} = \sum_i M_i \, r_i \, v_{t,i} $, where $r_i$ and $v_{t,i}$ is the distance to the center of mass and tangential velocity ($v_{t,i}=v_{k,i}\sin\alpha$, where $v_{k,i}$ is the kick velocity, and $\alpha$ is the angle between the velocity vector and radius vector). The vectors are in the ($x$, $y$,  $z$) coordinate system, of which the origin is the cluster center and the initial rotation of the cluster is oriented along the $z$ direction. For the analysis, we bin the escapers into consecutive time intervals of of 2.5\,Myr (until 10\,\text{Myr}), 10\,Myr (until 100\,\text{Myr}), and 30\,\text{Myr} (until dissolution; e.g., $0-2.5$\,\text{Myr}, $2.5-5$ \,\text{Myr}). Each escaping star was assigned to its respective interval based on its escape time, and all necessary mean and total values were calculated accordingly. After 10\,\text{Myr}, there are significant changes in the properties of the escapers. Firstly, the escape rate increases, which is accompanied by increases in $M_{\rm esc}$, $\overline{v}_{k}$, and $L_{\rm esc}$) starting from 20\,\text{Myr}. This occurs in all models, with the model with $\omega_0=1.8$ having the largest number of escapers. The escapers at this time show high $M_{\rm esc}$ and $\overline{v}_k$ regardless of initial rotation. Extreme values are seen in their angular momentum that reaches over $L_{\rm esc}=10^3$\,\text{$M_{\odot} \text{pc}^2 \text{Myr}^{-1}$} or $L_{\rm esc}=10^{-3}$\,\text{$M_{\odot} \text{pc}^2 \text{Myr}^{-1}$} with $50-300$ escaping stars. By comparing it to the bottom panel of Fig.~\ref{fig:groups} and considering the $L_{\rm esc}$ values, it is seen that those are influenced by the $M_{\rm mm}$ and $M_{\rm hm}$  stars. On further evolution, the parameter values of escapers show a stable decreasing rate for models with $\omega_0=[1.2;1.8]$ and an increasing rate for models with $\omega_0=[0.0;0.6]$ until dissolution.

\subsubsection{Change of rotation direction}
\label{sec:rot_direct}

\begin{figure}[tb]
    \centering
    \includegraphics[width=0.75\linewidth]{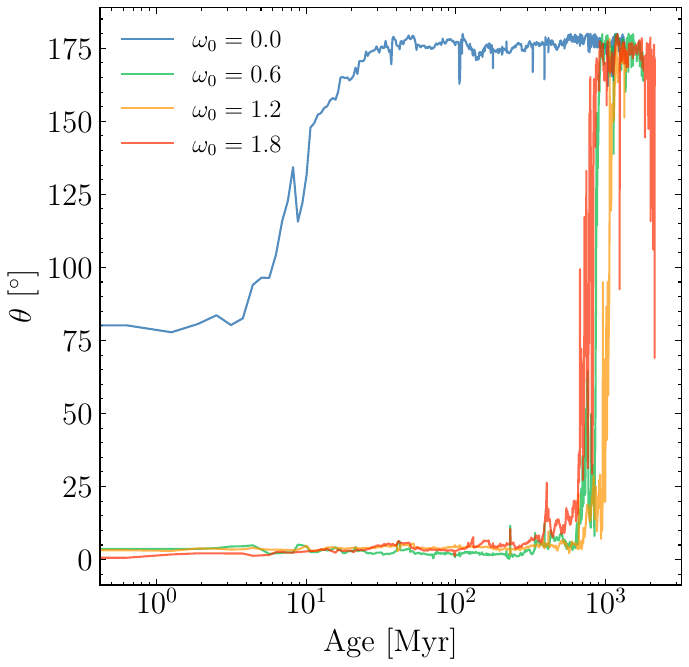}
    \caption{Angle, $\theta$, between the angular momentum vector and the $z$-axis as a function of the time of the models with 15k stars, $f_{\rm bin}=50\%$, and $Z=0.01$. The color scheme is the same as in Fig.~\ref{fig:ang_mom}. }
    \label{fig:phi}
\end{figure}

We also examine the change in the angle, $\theta$, between the cluster's net angular momentum vector and the $z$-axis of the Galactic plane. Fig.~\ref{fig:phi} shows the angle, $\theta$, as a function of time for the models with 15k stars, $f_{\rm bin}=50\%$, and $Z=0.01$. 

The rotating models ($\omega_0>0.0$) start with an angle very close to $0^\circ$, since the rotation axis is initialized along the $z$-axis. The evolution continues at a similar rate in all the rotating models, but the model with $\omega_0=1.8$ experiences moderate changes to $\theta\approx25^\circ$ at 400\,Myr. This is followed by the increase of angle $\theta$ until it reaches $\approx175^\circ$ by the age of $0.9-1$\,Gyr. The evolution in models with $\omega_0=[0.6,1.2]$ proceeds similarly, but the changes occur later compared to a model with $\omega_0=1.8$. Such a change is the indication that the models keep rotating along the $z$-axis, but in the opposite direction. This is the effect of the gravitational tidal field and the low number of stars in the system. These results explain the negative angular momentum seen in Fig.~\ref{fig:ang_mom}. 

The value of $\theta$ is not relevant for the nonrotating model. Even though the $\theta$ for the nonrotating model ($\omega_0=0.0$) starts at $\theta=38^\circ$, its total angular momentum is induced by the random motions of stars, which gives an arbitrary angle higher than $0^\circ$. The value of $\theta$ significantly increases after 10\,Myr, ultimately reaching $\theta\approx175^\circ$ by 100\,Myr. 

General minor changes in $\theta$ in rotating models are somewhat consistent with the previous work of \cite{tiongco2018}, but the change to the opposite side of the rotation axis is not seen there. The possible reason is that their simulations end earlier (upon reaching the $75\%$ mass loss) compared to the duration of our simulations (see Fig.~\ref{fig:4params}). Also, their simulations have 32,768 particles with equal mass.

\subsection{Dynamical evolution of binary systems and single stars}
\label{sec:bin}

In this section, we investigate the impact of the binary fraction, metallicity, and initial bulk rotation on the dynamical evolution of star clusters. We examine the change in global parameters and the rate of change of the number of objects present in the system.

\begin{figure}[tb]
    \centering
    \includegraphics[width=0.35\textwidth]{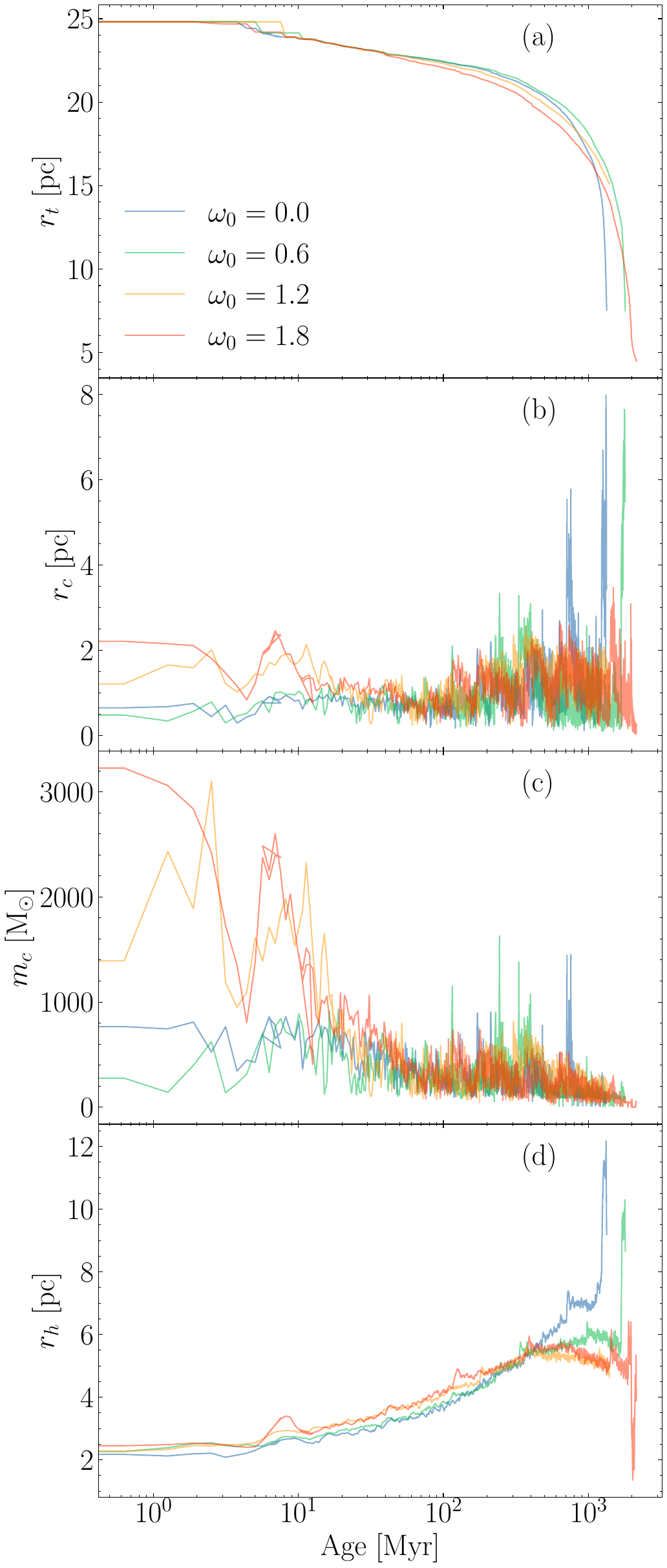}
    \caption{Evolution of several star cluster parameters, tidal radius $r_{\rm t}$ (panel (a)), the core radius $r_c$ (panel (b)), the core mass $m_c$ (panel (c)) and the half-mass radius $r_h$ (panel (d)) for simulations with 15k stars, $f_{\rm bin}=50\%$, $Z=0.01$. Different colors represent simulations with different $\omega_0$.}
    \label{fig:4params}
\end{figure}

Figure~\ref{fig:4params} shows the evolution of cluster parameters such as the tidal radius $r_{\rm t}$ (panel (a)), the core radius $r_c$ (panel (b)), the mass of the core $m_c$ (panel (c)), and the half-mass radius $r_h$ (panel (d)) in four panels from top to bottom. The core radius is the fundamental global parameter used for the characterization of the central structure of star clusters. The core mass is the mass of the stars within the core radius. The example models are the 4 models with 15k stars, $f_{\rm bin}=50\%$, $Z=0.01$ and different $\omega_0$. Colors represent the bulk rotation of the models. 

First, there is the time evolution of $r_{\rm t}$ in panel (a) of Fig.~\ref{fig:4params}. $r_{\rm t}$ decreases over time, and it is seen that the models with $\omega_0={1.2, 1.8}$ show a faster decrease over time compared to models with $\omega_0=[0.0, 0.6]$. However, during further evolution, models with $\omega_0=[0.0, 0.6]$ reach and even surpass the rate of change of $r_{\rm t}$ of models with high rotation. This leads to earlier cluster dissolution. 

\begin{figure*}
    \centering
    \includegraphics[width=\linewidth]{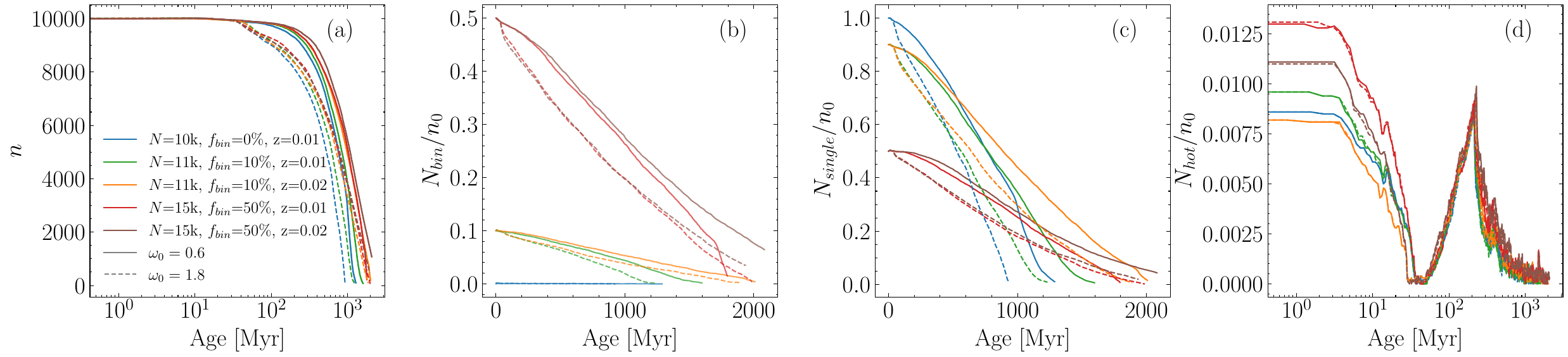}
    \caption{Number of single stars and binary systems  ($n=N_{\rm single}+N_{\rm bin}$; panel (a)), the fraction of remaining binary systems ($N_{\rm bin}/n_0$; panel(b)), single stars ($N_{\rm single}/n_0$; panel (c)), and hot stars with $T_{\rm eff}>20\,000$\,K ($N_{\rm hot}/n_0$; panel (d)) as a function of time for the models with $\omega_0=0.6$ and $\omega_0=1.8$. The term $n_0$ is the initial number of single stars and binary systems. $N_{\rm bin}$, $N_{\rm single}$, $N_{\rm hot}$ are the number of remaining binary systems, single stars, and hot stars. The colors represent models with different $N$ (total initial number of stars), $f_{\rm bin}$ (initial binary fraction) and $Z$ (metallicity). Line styles represent simulations with different $\omega_0$.  }
    \label{fig:hot_b_ratio}
\end{figure*}

Panels (b), (c), and (d) of Fig.~\ref{fig:4params} show the time evolution of the evolution of $r_c$, $m_c$, and $r_h$. Initially, $r_c$ and $m_c$ are higher among the models with greater rotation. Further, at the age of $\approx2-5$\,\text{Myr}, there is a huge decrease in $r_c$ and $m_c$ among all the models. This is the core collapse. Noticeably, the models with higher rotation experience much stronger core collapse. At the age of $5-10$\,Myr, there is another increase of $r_c$ and $m_c$, especially seen in the models with high rotation. The second core collapse further follows this at $\approx10-15$\,Myr in the models with high rotation. The early evolution of $r_h$ up to 400\,Myr is similar in all the models. The higher $r_h$ is seen in models with high rotation. One particular increase is seen at $\approx7-10\,\text{Myr}$ in the model with $\omega_0=1.8$. This time corresponds to the increase in $r_c$ and $m_c$. Further in the evolution, the clusters with higher bulk rotation have decreased $r_h$. This is due to the high mass loss of these models (see Fig.~\ref{fig:hot_b_ratio}).

It is known from previous studies that binary stars act as gravitational energy sources  \citep{aarseth1985, heggie1975, heggie1984, henon1975, trenti2007}. The dynamical formation of binaries primarily occurs near the center of the cluster \citep{bettwieser&sugimoto1984}. Additionally, binary stars should supposedly halt the contraction, affecting the gravogyro catastrophe \citep{aarseth1985, heggie1975, heggie1984}. The high binary fraction can also delay or prevent the core collapse \citep{kamlah22}. Additionally, clusters with high primordial binary fractions have high gravitational binding energy, which means a higher chance of retaining stars and a longer cluster lifetime. For all these reasons, we observe the impact of primordial binaries on rotating clusters.

We investigate the rate of change of the number of objects. Fig.~\ref{fig:hot_b_ratio} shows the number of stars and binary systems  ($n=N_{\rm single}+N_{\rm bin}$; panel (a)), the fraction of remaining binary systems ($N_{\rm bin}/n_0$; panel(b)), single stars ($N_{\rm single}/n_0$; panel (c)), and hot stars with $T_{\rm eff}>20\,000$\,K ($N_{\rm hot}/n_0$; panel (d)) as a function of time for the models with $\omega_0=0.6$ and $\omega_0=1.8$. $n_0$ is the sum of the initial number of single stars and the initial number of binary systems. Finally, we define stars with $T_{\rm eff}>20\,000$\,K as hot stars.

From the rate of change in the total number of particles (panel (a)), binary systems (panel (b)), and single stars (panel (c)) of Fig.~\ref{fig:hot_b_ratio}, It is seen that the models with high $f_{\rm bin}$ and solar metallicity survive the longest. The models with similar $Z$ show almost the same rate of change of binaries and single stars. Nonetheless, models with $\omega_0=1.8$ show a decrease in the number of particles starting at $40\,\text{Myr}$, while models with $\omega_0=0.6$ start to decrease after $100\,\text{Myr}$. A decrease is seen in both single stars and binary systems. In the models with faster rotation $\omega_0=1.8$ (solid curves), the fraction of both binaries and singles decreases quicker than the models with slower rotation $\omega_0=0.6$ (dashed curves).

The hot stars (panel (d) of Fig.~\ref{fig:hot_b_ratio}), are stars with $T_{\rm eff}>20\,000$\,K. Models with an identical particle number but different abundance $Z$ have a similar number of massive stars. The model with 15k stars, $f_{\rm bin}=50\%$, and $Z=0.01$ has the highest number of hot stars, and the model with 11k stars, $f_{\rm bin}=10\%$, and $Z=0.02$ has the lowest amount of hot stars. Models with higher $Z$ tend to have fewer hot stars compared to the clusters with low $Z$. The number of hot stars is higher for models with a higher initial binary fraction.  For instance, the initial ratio of hot binary stars is more than 65\% for models with $N=15k$, $f_{\rm bin}=50\%$, and both $Z$ options. It is seen that the number of hot stars reaches its lowest value at around $30-40\,\text{Myr}$ for all models. This is followed by an increase of up to $200\,\text{Myr}$ and a constant decrease until the end of the dynamic evolution. The hot stars are mostly the most massive stars, and depletion by $30-40\,\text{Myr}$ can be explained by the short lifetime of massive stars and binary evolution. After this, they evolve into compact objects, such as neutron stars. The second peak of hot stars occurs due to the formation of carbon/oxygen WDs (COWD) and oxygen/neon (ONWD) WDs.

\subsection{3D morphology evolution}
\label{sec:morphology}

In this section, we investigate the shape of the star clusters in the simulations and the observations and compare our results with the observational data of 90 clusters obtained from Gaia DR3 \citep{gaiadr3} with the membership from \citet{ pang2021a, pang2022a,  pang2024}.  We investigate the morphology by fitting an ellipsoid to the cluster's shape and use its principal axes: the major axis $a$, the intermediate axis $b$, and the minor axis $c$, to quantify the cluster's structural and morphological properties at the different times. The principal axis ratios of the intermediate-to-major axial ratio $b/a$, the minor-to-major axial ratio $c/a$, and the minor-to-intermediate axial ratio $c/b$ characterize the degree of elongation and ellipticity of the cluster towards the $y$ and $z$ directions. We do note that we align $a, b, c$ axes to $x, y, z$ axes. This would show the development of morphological structures in the $x-y$ plane while $z$ is the rotational axis. 

\begin{figure}[tb]
    \centering
    \includegraphics[width=\columnwidth]{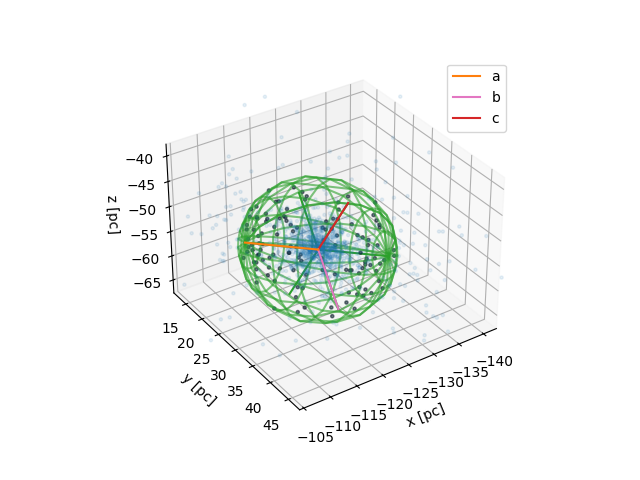}
    \caption{Ellipsoid drawn for the Pleiades cluster. The ellipsoid is represented by the green wireframe. Blue and black scatter points are the member stars and input stars of Pleiades used for the ellipsoid fitting. Orange, pink, and red lines indicate the principal axes $a$, $b$, and $c$. The stellar members of the Pleiades are obtained from \citet{pang2022a}. } 
    \label{fig:pleiades_example}
\end{figure}

The simulated clusters of \cite{kamlah22} develop a bar-like structure during the gravogyro catastrophe. To compare our results to those of \cite{kamlah22}, we fit the simulated cluster with a 3D ellipsoid model. The code of the ellipsoid fitting is adapted from the Ellipsoid-Fit code published on Github by R.\,M. Sandu\footnote{https://github.com/rmsandu/Ellipsoid-Fit.}. The fitted axis values depend on the number of stars that are included for the fitting. Stars in the inner part of the cluster do not contribute much to the ellipsoid fitting. Therefore, we only use the outer 10\% of stars within one tidal radius for the ellipsoid fitting. The ellipsoid fitting does measure the shape in the outer region of the cluster, which is influenced by the tidal field. However, the effects of internal rotation induce global changes in morphology, which also affect the outer part of the cluster, as can be seen from the decline of the axis ratio.  For illustration, we display the ellipsoid fitting of the Pleiades cluster, whose members are taken from \citet{pang2022a}, as an example in Fig.~\ref{fig:pleiades_example}. Blue and black scatter points are the member stars and the stars used for the ellipsoid fitting (outer 10\% of stars within the tidal radius), respectively. 

\begin{figure*}
    \centering
    \includegraphics[width=\textwidth]{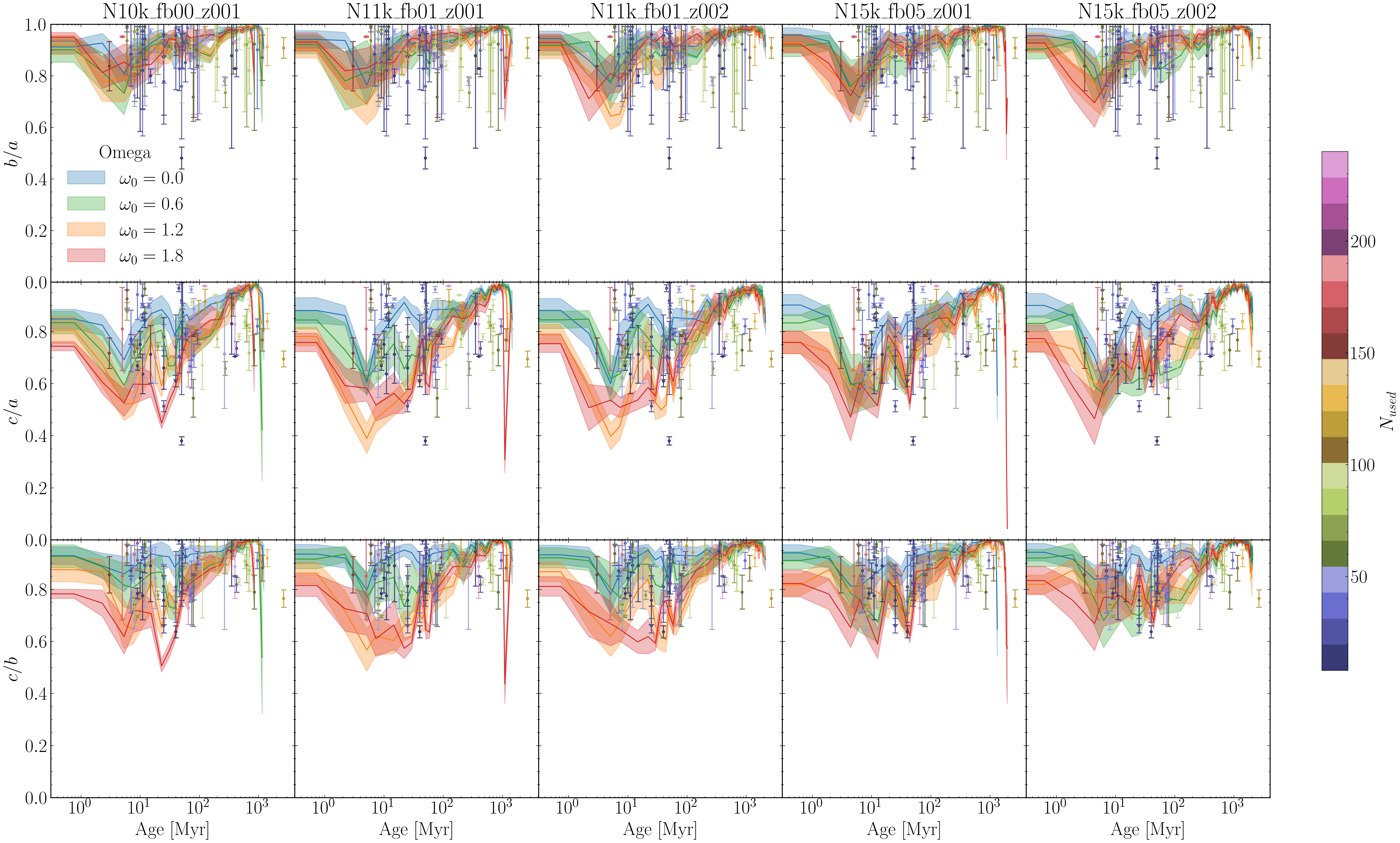}
    \caption{Evolution of principal axis ratios of the simulated clusters with age. The areas shaded in light colors enclose the minimum and maximum values from bootstrapping as a function of time, for intermediate-to-major axis ratio $b/a$ (upper row), the minor-to-major axis ratio $c/a$ (middle row), and the minor-to-intermediate axis ratio $c/b$ (bottom row). Columns show the results of the different models (indicated at the top). The colors of the shaded areas indicate the bootstrap range for models with different $\omega_0$. Symbols with error bars indicate the principal axis ratios for observations calculated from the members of the star clusters studied in \citet{ pang2021a,pang2022a,  pang2024}. }
    \label{fig:obs_abc_1tide}
\end{figure*}

Fig.~\ref{fig:obs_abc_1tide} presents the evolution of the principal axis ratios $b/a$ (upper row), $c/a$ ratios (middle row), and $c/b$ ratios (bottom row) of the simulated clusters and observations. The axis ratios of simulated clusters are the results of the bootstrapping procedure. We generate 100 bootstrap datasets\footnote{The bootstrap is the statistical data resampling technique that is used for the determination of various properties of the data \citep{bootstrap}.} from 1 snapshot and run ellipsoid fitting on each of the created datasets. We use mean values and standard deviation to construct the uncertainties in the principal axis ratios for the selected snapshot.  We choose 3 snapshots between 0 to 10 $N$-body time units (0, 3, and 7 $N$-body time), 9 between 10 to 100 $N$-body time units (at a 10 $N$-body time interval), and 9 between 100 to 1000 $N$-body time units (at a 100 $N$-body time interval). Models have different scaling factors of $N$-body units depending on $N$.  For the models with $N=15\,000$, one $N$-body time unit corresponds to a physical timescale of $\approx0.63$\,Myr. For $N=10\,000$ and $N=11\,000$, the scaling factors are $\approx0.76$ and $\approx0.73$, respectively.

Data points with error bars in Fig.~\ref{fig:obs_abc_1tide} are the observed clusters from \cite{pang2021a,pang2022a, pang2024}. Error bars are obtained by subtracting the difference between calculated principal axis ratios of clusters with and without corrected distances of \cite{pang2021a,pang2022a, pang2024}. The color pattern of observations is governed by the sample size used for the ellipsoid fitting. We note that the observation scatter points are identical in all panels.

Fig.~\ref{fig:obs_abc_1tide} shows that models have different axis ratios depending on the initial conditions (especially $\omega_0$) and on time. During the early evolution of the clusters, the largest differences in bootstrap ranges for models with different $\omega_0$ are present in the models with $N=10k$, $f_{\rm bin}=0\%$ and $Z=0.01$. These models have similar axis ratios from the start, ranging between $\approx85-94\%$ for $b/a$, $\approx73-84\%$ for $c/a$, and $\approx78-94\%$ for $c/b$. As by definition ($c\leq b \leq a$), all models have $c/a\leq b/a$. The model without rotation ($\omega_0=0.0$) starts with a more or less spherical morphology, with $c/b\approx c/a \approx b/a\approx 0.9\pm0.042$. The initial morphology in the models with stellar evolution from the work of \citet{kamlah22} did show similar triaxiality ( $b \neq a \neq c$ ) in the beginning.

Upon further evolution up until 10\,Myr, all the models experience a decrease in axis ratios, which is the effect of the core collapse. The models without bulk rotation also have a moderate decrease in the axis ratios, as they develop tidal tails under the influence of the Galactic tidal field. For the models with bulk rotation, the decrease is earlier and sharper, especially for the models with high bulk rotation. This is followed by differing increases and further decreases of axis ratios in all the models up until $100\,\text{Myr}$. The models without bulk rotation and $\omega_0=0.6$ have a increase in $b/a$,  $c/a$ and $c/b$ by the age of $20\,\text{Myr}$. The models with $\omega_0=1.2$ and $\omega_0=1.8$ show the same increase and decrease but with earlier peak and sharp decrease by the $20-25\,\text{Myr}$. This is especially apparent for the model with $\omega_0=1.8$ that reaches its all-time low in $c/a$. This kind of dynamic persists to $200\,\text{Myr}$. Clusters achieve high axis ratios until they are close to dissolution. The major change in $c/a$ indicates the existence of elongation and bar-like structure at the given moment. This kind of morphology was also apparent in the work of \citet{kamlah22}. Another similarity from \citet{kamlah22} is the change from initial triaxiality to axisymmetry ($b=a$, but $c\neq a $ and $c\neq  b$) in further evolution. The most drastic elongation reaching over $c/a\approx0.4$ can be seen for the model with $\omega_0=1.8$. This model has the highest angular momentum of the system and shows high ellipticity up until $50\,\text{Myr}$. This is consistent with the findings of \citep{lahen2020nov}, who investigated the structure and rotation of massive star clusters at $20\,\text{Myr}$. They found that for clusters with masses above 30\,000\,$M_{\odot}$, the higher the specific angular momentum, the higher the ellipticity (lower $c/a$).

All the other models with more particles and binaries perform similarly but have fluctuations at different timescales and with differing degrees. The second decrease in $c/a$, supposedly the second core collapse, is sharper and more apparent for the models without binaries. However, a similar decrease with a lesser scale is seen in the models with $Z=0.01$ as shown in the 2nd column and 4th columns from left in Fig.~\ref{fig:obs_abc_1tide}. The decrease is still present in models with $Z=0.02$, but it is even less apparent compared to models with $Z=0.01$. The decrease is sharpest for the models with $\omega_0=1.8$ with the only exception for the models with $N=11k$ and $f_{\rm bin}=10\%$. The model with $N=11k$, $f_{\rm bin}=10\%$, $Z=0.01$, and $\omega_0=1.2$ shows the sharpest decrease in both $b/a$ and $c/a$  at $4\,\text{Myr}$ among all the models.  

In terms of the observations, most of the selected observable clusters are inside the bootstrap ranges of the simulations. The ones that have higher axis ratios are the clusters with a low number of input stars for ellipsoid fitting. The exceptions are the NGC\,1980, NGC\,2516, and the Pleiades, which have more than 100 input samples and show both axis ratios above 0.9. The discrepancy of the observed clusters is $\approx40\%$. Non-conformant clusters majorly tend to have input samples of less than 100 stars. The only exception is the NGC\,2516, which has $b/a$, $c/a$ and $c/b$ of $<0.97$, despite having input stars of $\approx200$. Those that have lower $c/a$ compared to the simulation bootstrap ranges are mostly filamentary clusters such as Collinder\,132\,gp3, Collinder\,132\,gp5, and NGC\,6405. They are young and elongated, thus showing $c/a<0.6$. They already formed the tidal tails and were considered dynamically older compared to the simulations at a similar age.

\subsection{Synthetic observations for simulated clusters}
\label{sec:galevnb}

Here, we present our simulations in three observational magnitude filters and spectral energy distribution (SED). We used \texttt{GalevNB} \citep{pang2016galev} to convert the physical properties of $N$-body simulation data into filter bands of Gaia, China Space Station Telescope (CSST), and Hubble Space Telescope/ Space Telescope Imaging Spectrograph  (HST/STIS).  This allows for direct comparison between numerical studies and observations. The SED is needed to investigate excess ultraviolet (UV) emission. We converted the data into spectra spanning far-UV to near-IR wavelengths. The SED was created by summing the fluxes of the stellar population. We further separated the summation of fluxes into two categories: single stars and binary systems. 

\begin{figure*}[tb]
    \centering
    \includegraphics[width=0.95\textwidth]{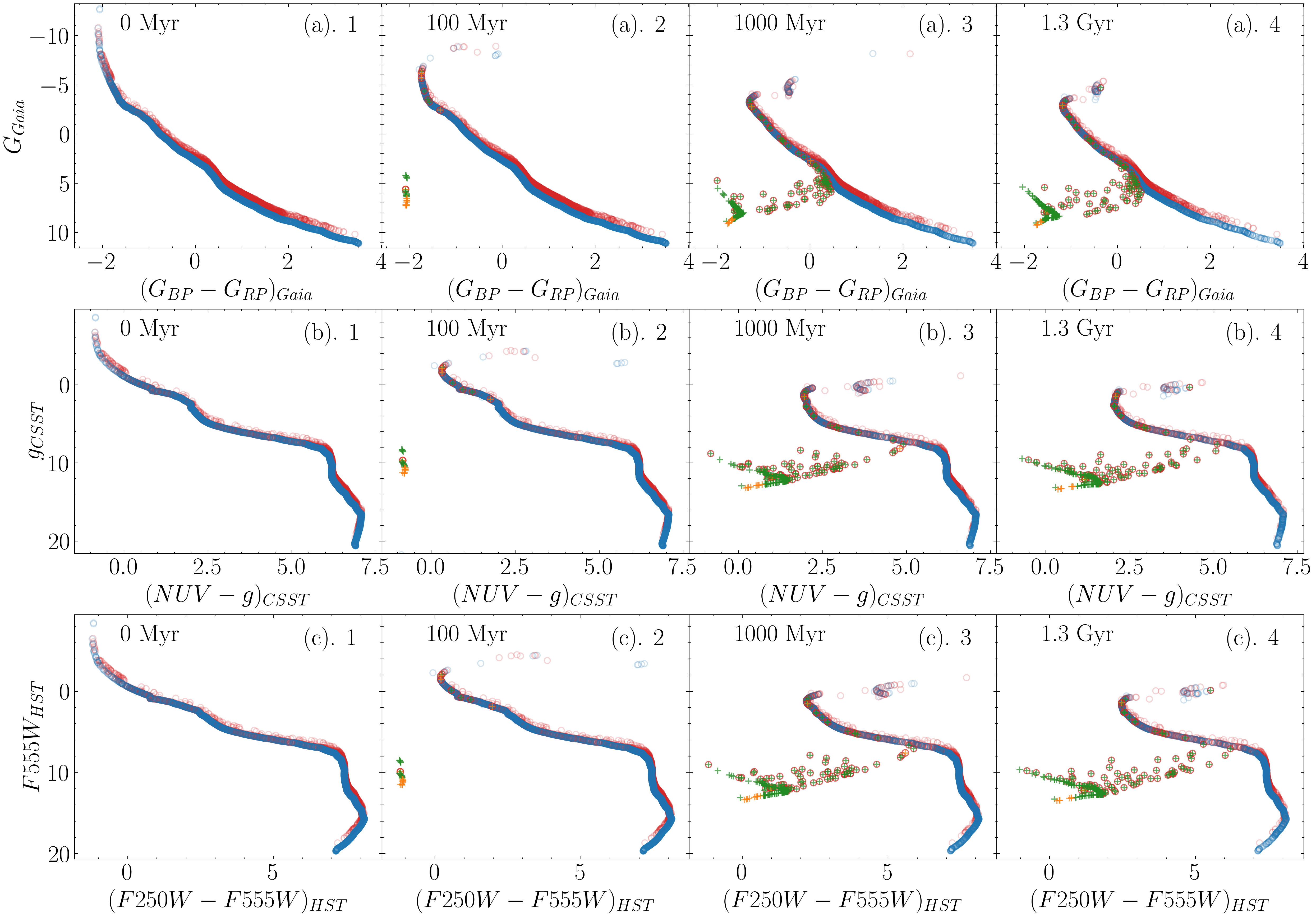}
    \caption{Color-magnitude diagrams for three filter bands at ages between 0\,Myr to 1.3\,Gyr, for the cluster model with 15k stars, $f_{\rm bin}=50\%$, $Z=0.01$, and $\omega_0=1.2$. Rows represent CMDs for different filters. Columns represent different ages. Blue and red circles represent the magnitude of single stars and binary systems. Green and orange crosses inside the circles are the COWDs and ONWDs.} 
   \label{fig:cmd_galev}
\end{figure*}

Fig.~\ref{fig:cmd_galev} shows the CMDs of the simulated clusters in three observational filters for ages between $0\,\text{Myr}$ to $1.3\,\text{Gyr}$ for the model with 15k stars, $f_{\rm bin}=50\%$, $Z=0.01$, and $\omega_0=1.2$. As can be seen from the figure, the magnitude ranges of CSST and HST are very similar to that of Gaia. We note that for a binary to be classified as a WD binary, at least one of the two components should be a WD.  Initially, the majority of the single stars and binaries are on the main sequence. The number of binary systems is equal to the number of single stars. After 100\,Myr, there are ONWDs that are mostly single stars. However, in further evolution, there are mostly COWDs. Single-star WDs formed the WD sequence on the bottom left side of the CMD. Binaries with WDs are either near the main sequence or between the main sequence and the WD sequence. The ones between them are the cataclysmic variable star binaries, which are interacting WD binaries that have an MS star companion with a mass comparable to the WD \citep{warner1995, pang2022cv}.  

\begin{figure*}[tb]
    \centering
    \includegraphics[width=0.95\textwidth]{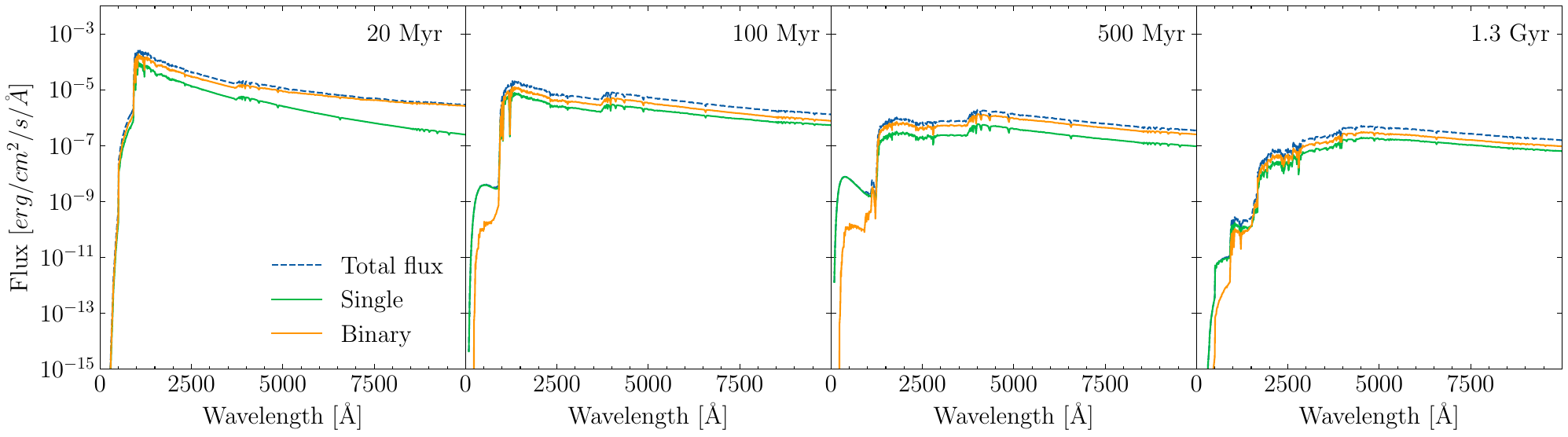}
    \caption{Spectral energy distributions for ages between 0\,Myr to 1.3\,Gyr, for the model with 15k stars, $f_{\rm bin}=50\%$, $Z=0.01$, and $\omega_0=1.2$. The colors of the curves represent the total flux, single star flux, and binary system flux as dashed blue, green, and orange lines respectively. }
   \label{fig:sed_galev}
\end{figure*}

Fig.~\ref{fig:sed_galev} shows SEDs for ages from 0\,Myr to 1.3\,Gyr for the model with 15k stars, $f_{\rm bin}=50\%$, $Z=0.01$, and $\omega_0=1.2$. The curves represent the total flux (blue), single star flux (green), and binary system flux (orange). In the models with $f_{\rm bin}=50\%$, binaries have twice the flux the of the single stars.

\section{Discussion}
\label{sec:dis}

In this section, we discuss and interpret our findings by examining several parameters simultaneously. Furthermore, we compare our results with findings from the literature. 

\begin{figure*}[tb]
    \centering
    \includegraphics[width=0.9\textwidth]{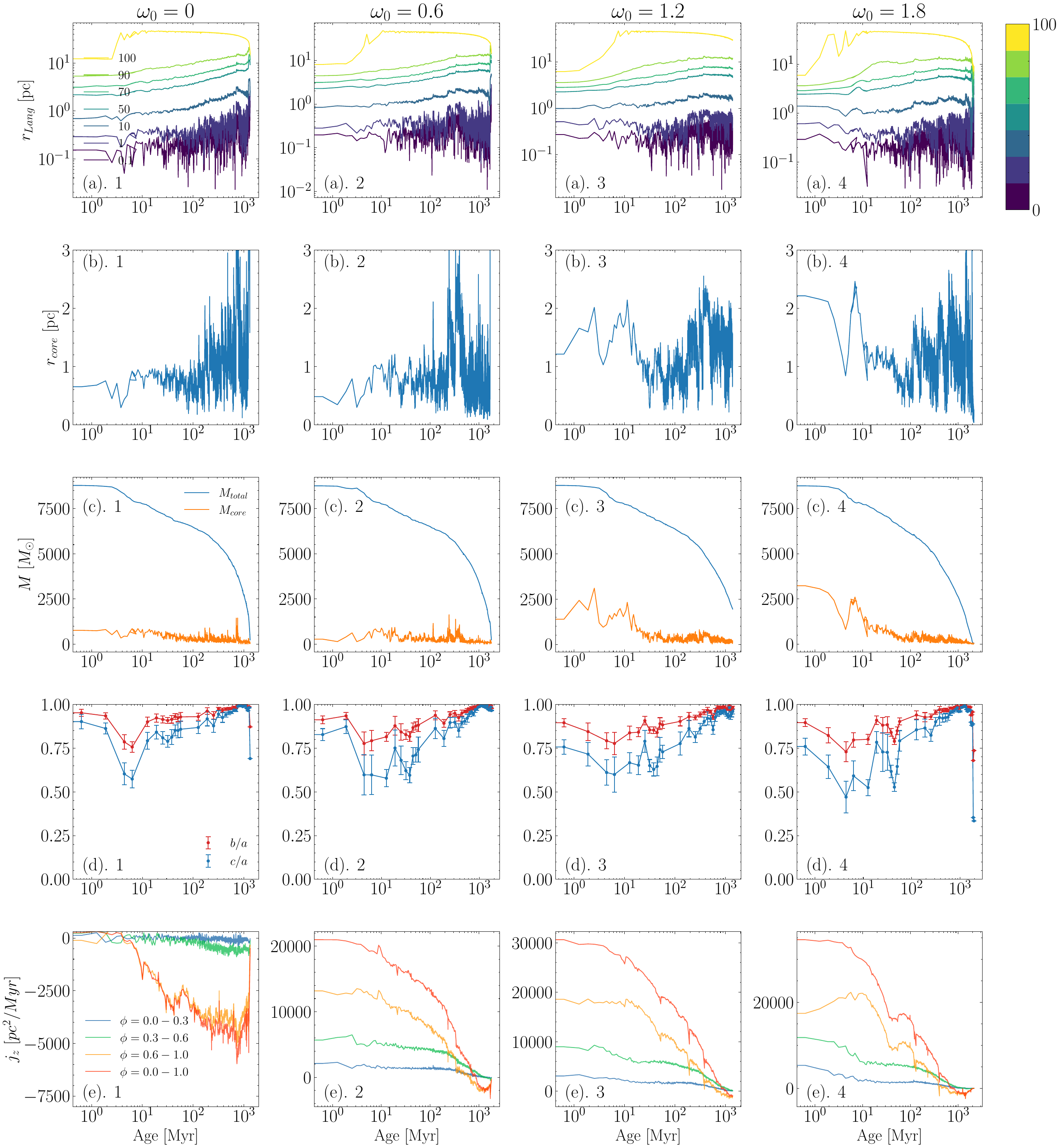}
    \caption{Lagrangian radii $r_{\rm lagr}$ (panels (a)), core radii $r_{\rm core}$ (panels (b)), total mass $M_{\rm total}$ and core mass $M_{\rm core}$ (panels (c)), principal axes ratios $b/a$ and $c/a$ (panels (d)), and angular momentum on the $z$-axis (panels (e)) of the cluster model with 15k stars, $f_{\rm bin}=50\%$, $Z=0.01$, and different $\omega_0$. Rows represent respective parameters, and columns represent the initial degree of rotation.  The upper row shows Lagrangian radii, where color map colors correspond to the radius of different mass percentages. The colors in the 4th row show the principal axis for the stars within one tidal radius. Colors in the bottom row show the angular momentum in $z$-axis for all of the stars within the inmost part ($\phi=0.0-0.3$), central part ($\phi=0.3-0.6$), outmost part ($\phi=0.6-1.0$), and the entire system ($\phi=0.0-0.1$). 
    }
    \label{fig:all}
\end{figure*}

Fig.~\ref{fig:all} shows a comparison of the Lagrangian radii $r_{\rm lagr}$ (panels (a)), core radii $r_{\rm core}$ (panels (b)), total mass $M_{\rm total}$ and core mass $M_{\rm core}$ (panels (c)), principal axes ratios $b/a$ and $c/a$ (panels (d)), and angular momentum on the $z$-axis (panels (e)) of the model with 15k stars, $f_{\rm bin}=50\%$, $Z=0.01$, for different values of $\omega_0$.

The Lagrangian radii (Fig.~\ref{fig:all}; top row) show that 90\% of the cluster mass remains within a similar distance with a slight expansion over the course of the simulation. The radii that change a lot are those with $\phi=0.9$ and $\phi=1.0$. Notably, the $\phi=1.0$ radius increases drastically, and then remains more or less constant until dissolution. The expansion of this part starts as early as 3\,Myr, but it is seen that it occurs sooner for the models with high $\omega_0$. For example, for the model with $\omega_0=1.8$ it starts expanding at 1\,Myr and reaches its peak of $\approx40$\,pc around $\approx2.5-3$\,Myr. 

Fig.~\ref{fig:all} shows the evolution of $r_{\rm core}$ (row (b)), and $M_{\rm total}$ and $M_{\rm core}$ (row (c)). These panels allowed us to identify core collapses, as well as the changes in total cluster mass. A significant downward change in $r_{\rm core}$ and $M_{\rm core}$ occurs in all models. The change is greater in the models with $\omega_0=[1.2, 1.8]$. Simultaneously, $M_{\rm total}$ decreases significantly. It is seen that rapidly rotating models experience additional core collapses after 50\,Myr. Another notable feature is that models with higher rotation have a higher core mass and core radius at the start of the simulations.  This is consistent with the hydrodynamic simulations, where it was seen that the rotation signature is more apparent for the high-mass clusters \citep{mapelli2017}.

The effects of core collapse are also seen in the principal axes ratio evolution shown in row (d) of Fig.~\ref{fig:all}. At the same timescales of the core collapse shown in row (b), there are major changes in $c/a$, showing the elongation among the models. The higher the bulk rotation, the more apparent the elongation at the exact time periods of angular momentum transport. Additionally, rapidly rotating models have another significant elongation at the timescale of the second core collapse. Elongation at this time is more prominent in the models without primordial binaries (see Section~\ref{sec:morphology}).

The effects of core collapse are seen in the angular momentum shown in the bottom row (e) of Fig.~\ref{fig:all}. Colors in row~5 represent angular momenta at different Lagrangian radii, where blue, green, yellow, and red correspond to angular momentum in the $z$-axis for all of the stars, the inmost part ($\phi=0.0-0.3$), central part ($\phi=0.3-0.6$), outmost part ($\phi=0.6-1.0$), and the entire system ($\phi=0.0-1.0$). Prior to the first core collapse, there is angular momentum transport from the inner part and the central parts to the outer regions of the clusters. This is more apparent in the model with $\omega_0=1.8$. Further, the momentum of the outer part decreases until 50\,Myr, where there is another increase that leads to the second core collapse. The consequences of this process can also be observed in other parameters, such as $r_{\rm core}$ and $c/a$.

All this shows the effects of the gravothermal-gravogyro catastrophe on the dynamic evolution of the star clusters. The gravogyro catastrophe can be noticed by the angular momentum transport. The culmination of the gravothermal-gravogyro catastrophe is the 1st core collapse that happens in every observed model. If the cluster happens to have high bulk rotation, then there might be a second core collapse that can be seen through the instability in angular momentum and 3D morphology. If the observed clusters also happen to have primordial binaries, it can somewhat hold out the second core collapse.

\section{Conclusion}
\label{sec:conc}

We examined the dynamical evolution of the rotating star clusters with differing primordial binary fractions, as well as different metallicities. We examined the evolution of the rotation signature, the impact of binaries and single stars, and the 3D morphology of the rotating star clusters. We also converted our simulations to the current observational filters for future comparison with observations. Our main findings can be summarized as follows:

\begin{enumerate}

    \item The models with a high bulk rotation experience two core collapses before the age of 20\,Myr. This is due to the gravogyro-gravothermal catastrophes seen by angular momentum transport from the inner parts to the outer. There is an increase in the number of escapers at this period, and they are induced from $M_{\rm mm}$ and $M_{\rm hm}$ stars under stellar evolution. 
    
    \item The effect of bulk rotation is noticeable during the early phases of evolution, but disappears by the timescale of two-body relaxation. 
    
    \item There is a change in the direction of the angular momentum vector, which occurs early in the evolution for models with no rotation, and near the time of dissolution for models with rotation. This change is caused by the tidal field. The effects are stronger in nonrotating clusters, and rotating clusters are affected after the disappearance of the rotation signature and eventual mass loss. 
    
    \item Bulk rotation affects the long-term dynamic evolution and strongly affects the morphology at the early stages of evolution. The effect is seen in high elongation at the timescales of angular momentum transport. 
    
    \item We presented synthetic observations of the simulated clusters for future comparison with observations. These results can be used for the interpretation of future observations of Gaia, CSST, and HST.
    
\end{enumerate}

Our study has several limitations that will be addressed in future studies. We have analyzed simulations with $\sim 10^4$ stars. We plan to extend our study with simulations of 50\,000 particles to assess the effects of rotation and primordial binaries on more massive stellar systems. We will examine their dynamic evolution and attempt to confirm the relationship between the size of the cluster and the ellipticity and angular momentum transport described in \cite{lahen2020nov}. We will similarly examine the dynamical evolution and convert the simulation results to different observational filter bands. 


\section*{Data availability}

The snapshot data of the models used in this study are available at Zenodo, \url{https://zenodo.org/records/15637954}. Uploaded snapshot files have basic data of single stars, binaries, and mergers at different ages, with an interval of 10 Myr before the age of 100 Myr. After 100 Myr until the end of simulations, the interval of snapshots is increased to 100 Myr. Refer to README file on Zenodo for instructions. 

If more snapshot data or other output data is required, it can be requested from the corresponding author. 

\begin{acknowledgements}
We thank the referee for the constructive comments that helped us to improve this paper in many ways. We thank Dr. Michela Mapelli and Dr. Jiadong Li for helpful discussions. 
Xiaoying Pang acknowledges the financial support of the National Natural Science Foundation of China through grants 12173029 and 12233013. We acknowledge the science research grants from the China Manned Space Project with No.~CMS-CSST-2021-A08. 
This research is supported by the Science Committee of the Ministry of Science and Higher Education of the Republic of Kazakhstan (Grant No.~AP19677351 and AP13067834).  
R.S. acknowledges NAOC International Cooperation Office for its support in 2023, 2024, and 2025; and the National Science Foundation of China (NSFC) under grant No. 12473017. This research was supported in part by grant NSF PHY-2309135 to the Kavli Institute for Theoretical Physics (KITP). R.S. acknowledges support by the German Science Foundation (DFG, project Sp 345/24-1).
M.G. was supported by the Polish National Science Center (NCN) through the grant 2021/41/B/ST9/01191.
A.A. acknowledges support for this paper from project No.~2021/43/P/ST9/03167 co-funded by the Polish National Science Center (NCN) and the European Union Framework Programme for Research and Innovation Horizon 2020 under the Marie Skłodowska-Curie grant agreement No.~945339. 
P.B. thanks the support from the special program of the Polish Academy of Sciences and the U.S. National Academy of Sciences under the Long-term program to support Ukrainian research teams grant No.~PAN.BFB.S.BWZ.329.022.2023. P.B. also thanks the support under the project No.~BR24992759 ``Development of the concept for the first Kazakhstani orbital cislunar telescope - Phase I'', financed by the Ministry of Science and Higher Education of the Republic of Kazakhstan.

Software:
{\texttt{Astropy} \citep{astropy2013,astropy2018,astropy2022}, 
            \texttt{SciPy} \citep{millman2011}, 
            \texttt{NBODY6++GPU} \citep{kamlahnbody},
            \texttt{McLuster} \citep{kupper2011, kamlahnbody}, and
            \textsc{FOPAX} \citep{einsel1999spurzem, kim2002einsel, kim2004lee, kim2008yoon}.
}
\end{acknowledgements}



\bibliography{aa54093-25}
\bibliographystyle{aasjournal}

\end{document}